\documentclass[conference]{IEEEtran}

\usepackage{cite}
\usepackage{graphicx}
\usepackage{subcaption}
\usepackage{xcolor}
\usepackage{tikz}
\usepackage{pgfplots}
\usepackage{multirow}
\usepackage{upgreek}
\usepackage{amssymb}
\usepackage{amsmath}
\usepackage{cases}
\usepackage{pifont}
\usepackage{bm}
\usepackage{amsthm}

\usepackage{tabularx}
\usepackage{booktabs}
\newcolumntype{Y}{>{\centering\arraybackslash}X}
\usepackage{array}
\usetikzlibrary{arrows,positioning,shapes.geometric,spy}
\usepackage{float}
\usepackage{algorithmic}
\usepackage{xcolor}
\usepackage[linesnumbered,ruled,vlined]{algorithm2e}
\usepackage{changes}
\usepackage{bbm}
\usepackage{url}

\newcommand{\fixme}[2]{\ifx&#2&{\leavevmode\color{red}#1}\else{\leavevmode\color{red}FIXME\{}#1{\leavevmode\color{red}\}}\footnote{{\leavevmode\color{red}#2}}\PackageWarning{Fixme}{#1: #2}\fi}

\SetCommentSty{mycommfont}

\DeclareMathOperator*{\argmax}{arg\,max}
\DeclareMathOperator*{\sgn}{sgn}
\DeclareMathOperator{\PM}{PM}
\DeclareMathOperator*{\arctanh}{arctanh}

\begin{document}

\title{Decoding Reed-Muller and Polar Codes by Successive Factor Graph Permutations}

\author{\IEEEauthorblockN{Seyyed Ali Hashemi\IEEEauthorrefmark{1}, Nghia Doan\IEEEauthorrefmark{1}, Marco Mondelli\IEEEauthorrefmark{2}, Warren J. Gross\IEEEauthorrefmark{1}}
\IEEEauthorblockA{
\IEEEauthorrefmark{1}Department of Electrical and Computer Engineering, McGill University, Montr\'eal, Qu\'ebec, Canada\\
\IEEEauthorrefmark{2}Department of Electrical Engineering, Stanford University, California, USA\\
Email: seyyed.hashemi@mail.mcgill.ca, nghia.doan@mail.mcgill.ca, mondelli@stanford.edu, warren.gross@mcgill.ca}}

\maketitle
\begin{abstract}
Reed-Muller (RM) and polar codes are a class of capacity-achieving channel coding schemes with the same factor graph representation.
Low-complexity decoding algorithms fall short in providing a good error-correction performance for RM and polar codes. Using the symmetric group of RM and polar codes, the specific decoding algorithm can be carried out on multiple permutations of the factor graph to boost the error-correction performance. However, this approach results in high decoding complexity. In this paper, we first derive the total number of factor graph permutations on which the decoding can be performed. We further propose a successive permutation (SP) scheme which finds the permutations on the fly, thus the decoding always progresses on a single factor graph permutation. We show that SP can be used to improve the error-correction performance of RM and polar codes under successive-cancellation (SC) and SC list (SCL) decoding, while keeping the memory requirements of the decoders unaltered. Our results for RM and polar codes of length $128$ and rate $0.5$ show that when SP is used and at a target frame error rate of $10^{-4}$, up to $0.5$~dB and $0.1$~dB improvement can be achieved for RM and polar codes respectively.
\end{abstract}


\IEEEpeerreviewmaketitle

\section{Introduction} \label{sec:intro}

The next generation of wireless communications standard (5G) requires channel coding techniques for the control channel which are of short length. Polar codes \cite{arikan} are a class of channel coding techniques that were recently selected for the enhanced mobile broadband (eMBB) control channel of 5G \cite{3gpp_polar}. Initially, polar codes attracted a lot of research attention because of their capacity achieving property at infinite code length, with the low-complexity successive-cancellation (SC) decoding. However, for the 5G control channel applications with short code lengths, SC falls short in providing a reasonable error-correction performance.
A SC list (SCL) decoder can reduce the error-correction performance gap between SC and maximum a posteriori (MAP) decoder \cite{tal_list}. However, the error-correction performance of polar codes under MAP decoding is not satisfactory.

Reed-Muller (RM) codes \cite{Muller_RM,Reed_RM} are similar to polar codes in the sense that the generator matrices of both codes are constructed by selecting rows from a Hadamard matrix. The row selection of polar codes minimizes the error probability under SC decoding, while the row selection of RM codes maximizes the minimum distance. As a result, polar codes outperform RM codes under SC decoding and RM codes outperform polar codes under MAP decoding. Recently, it was shown that RM codes achieve the capacity of a binary erasure channel (BEC) under MAP decoding \cite{Kudekar_RM}. Since MAP decoding is practically intractable, sub-optimal decoding algorithms such as SC \cite{Dumer_SC} and SCL \cite{Dumer_SCL} are used to decode RM codes. However, SC decoding provides a poor error-correction performance when used to decode RM codes and SCL decoding requires a large list size to achieve a desirable error rate.

The symmetric group of RM and polar codes were used to improve their error-correction performance. In particular, a number of permutations on the factor graph representation of polar and RM codes is selected and the decoding algorithm is performed on them \cite{KEY_RMPerm,Kor09thesis}. It was shown in \cite{Dumer_SCL} that using the factor graph permutations of RM codes helps boost the error-correction performance of SCL decoding. Moreover, \cite{Kor09thesis} showed that for a polar code of length $N = 2^n$, there are $n!$ factor graph permutations on which the decoding can be performed. To limit the complexity of the decoder, it was suggested in \cite{Kor09thesis} and \cite{elkelesh2018belief} that only the cyclic shifts or random permutations are selected for decoding. However, the error-correction performance of decoding polar codes under the aforementioned schemes deteriorated significantly in comparison with using all the factor graph permutations. In \cite{Doan_GLOBECOM}, a method was derived to carefully select the factor graph permutations in order to improve the error-correction performance of polar codes. However, all of these techniques require multiple decoding attempts on different factor graph permutations, which adversely affects the decoder complexity.

In this paper, we first show that by considering the permutations on polar code partitions \cite{hashemi_TCOM}, the number of factor graph permutations is larger than $n!$. We then use these permutations to propose a successive permutation (SP) scheme which can find the suitable permutations during the decoding process on the fly. We show that this approach can significantly improve the performance of RM codes under SC decoding with low complexity overhead. We further apply the proposed SP scheme to the SCL decoder for RM codes and show that SP can help SCL decoder to approach MAP decoding performance with a smaller list size than the SCL decoder alone. We then devise a method to apply SP on SC and SCL decoding of polar codes to improve their error-correction performance. Since the SP scheme enables the decoder to progress on a single factor graph, the memory requirements of the underlying decoder remains unchanged. Our results show that for a code of length $128$ and rate $0.5$ at a target frame error rate (FER) of $10^{-4}$, SP results in up to $0.5$~dB improvement for SC and SCL decoding of RM codes, and up to $0.1$~dB improvement for SC and SCL decoding of polar codes.

\section{Preliminaries} \label{sec:prel}

\subsection{Construction of Polar and RM Codes}

A polar code of length $N = 2^n$ with $K$ information bits is denoted as $\mathcal{P}(N,K)$. Similarly, a RM code of length $N = 2^n$ and dimension $K$ is represented as $\mathcal{RM}(N,K)$. Let us denote by $R = \frac{K}{N}$ the rate of the code. The encoding process of polar and RM codes includes a matrix multiplication as
\begin{equation}
\mathbf{x} = \mathbf{u}\mathbf{G}^{\otimes n} \text{,} \label{eq:polarEnc}
\end{equation}
where $\mathbf{x} = \{x_0,x_1,\ldots,x_{N-1}\}$ is the sequence of coded bits, $\mathbf{u} = \{u_0,u_1,\ldots,u_{N-1}\}$ is the sequence of input bits, and $\mathbf{G}^{\otimes n}$ is the $n$-th Kronecker power of the matrix $\mathbf{G} = \left[\begin{smallmatrix} 1&0\\ 1&1 \end{smallmatrix} \right]$.

Construction of polar and RM codes relies on the selection of $K$ rows of matrix $\mathbf{G}^{\otimes n}$ to build the generator matrix. This process is equivalent to assigning information bits to the corresponding $K$ elements in $\mathbf{u}$ and setting the remaining $N-K$ bits to $0$. In this paper, we denote by the indices in the vector $\mathbf{u}$ which carry information as the information set $\mathcal{A}$ and we call the remaining bit indices as the frozen set $\mathcal{A}^c$.

In case of polar codes, the selection of information set $\mathcal{A}$ is performed by using the polarization property. It was shown in \cite{arikan} that as the code length goes to infinity, the bit-channels polarize in the sense that some of them become completely noisy and the others become completely noiseless. The fraction of noiseless bit-channels tends to the capacity of the channel. For finite practical code lengths, the bit-channels are divided into two sets $\mathcal{A}$ and $\mathcal{A}^c$ depending on the number of information bits. It should be noted that the polarization process is dependent on the channel over which the data transmission takes place.

In case of RM codes, the information set $\mathcal{A}$ is generated by selecting the rows of $\mathbf{G}^{\otimes n}$ with largest Hamming weights. For a RM code with minimum distance $2^m$, the rows with minimum Hamming weight $2^m$ are selected. Therefore, the code rate for RM codes is constrained to
\begin{equation}
R_{\text{RM}} = \frac{\displaystyle{\sum_{i = 0}^{m} {n \choose i}}}{N} \text{,} \label{eq:RMRate}
\end{equation}
where $0 \leq m \leq n$. The advantage of RM codes over polar codes is that the construction of RM codes is channel-independent.

After $\mathbf{u}$ is populated, the matrix multiplication in (\ref{eq:polarEnc}) is performed to generate $\mathbf{x}$. Subsequently, $\mathbf{x}$ is modulated and transmitted through the channel. In this paper, we consider binary phase-shift keying (BPSK) modulation and binary additive white Gaussian noise (AWGN) channel model with variance $\sigma^2$. Therefore, the received vector $\mathbf{y} = \{y_0,y_1,\ldots,y_{N-1}\}$ can be translated into logarithmic likelihood ratio (LLR) domain as $\bm{\alpha} = \frac{2}{\sigma^2} \mathbf{y}$, where $\bm{\alpha} = \{\alpha_0, \alpha_1, \ldots, \alpha_{N-1}\}$.

\subsection{Decoding of Polar and RM Codes}

\begin{figure}[t]
  \centering
  \label{fig:BPDec}	
  \begin{subfigure}[b]{0.5\textwidth}
  	\centering
  	\includegraphics[width=0.7\linewidth]{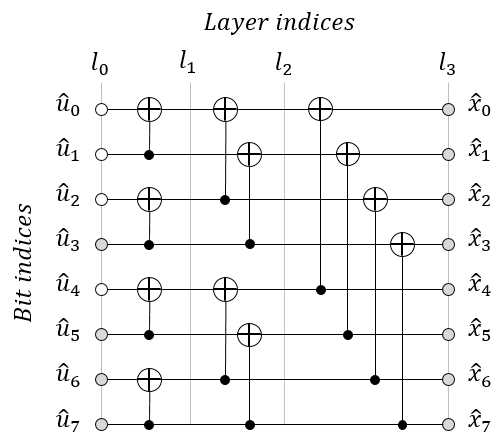}
  	\caption{}
  	\label{fig:BPDec:a}
  \end{subfigure}  
  \begin{subfigure}[b]{0.5\textwidth}
  	\centering
	\vspace*{0.2cm}
  	\hspace*{0.6cm}
  	\includegraphics[width=0.8\linewidth]{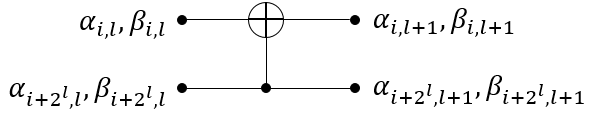}
  	\caption{}
  	\label{fig:BPDec:b}
  \end{subfigure}
  \caption{(a) Factor graph representation of a polar (RM) code with $N=8$, $K=4$, and $\{u_0,u_1,u_2,u_4\}\in \mathcal{A}^c$, (b) a processing element (PE).}
\end{figure}

Decoding of polar and RM codes can be represented on a factor graph as shown in Fig.~\ref{fig:BPDec:a} for a code of length $N=8$ with $K=4$. The messages are propagated through the processing elements (PEs) as shown in Fig.~\ref{fig:BPDec:b} where $\beta_{i,l}$ denotes a left-to-right message, and $\alpha_{i,l}$ denotes a right-to-left message of the $i$-th bit index at layer $l$ of the factor graph.

\subsubsection{Successive-Cancellation Decoding}

In SC decoding, the right-to-left messages $\alpha_{i,l}$ are soft LLR values and the left-to-right messages $\beta_{i,l}$ are hard decision bits. The messages in the factor graph are updated as
\begin{align}
\alpha_{i,l} &= f\left(\alpha_{i,l+1},\alpha_{i+2^l,l+1}\right) \text{,} \\  
\alpha_{i+2^l,l} &= g\left(\alpha_{i,l+1},\alpha_{i+2^l,l+1},\beta_{i,l}\right) \text{,}\\ 
\beta_{i,l+1} &= \beta_{i,l} \oplus \beta_{i+2^l,l} \text{,}\\
\beta_{i+2^l,l+1} &= \beta_{i+2^l,l} \text{,}
\label{eq:PEUpdateSC}
\end{align}
where
\begin{align}
f\left(a,b\right) &= 2\arctanh \left(\tanh\left(\frac{a}{2}\right)\tanh\left(\frac{b}{2}\right)\right) \nonumber \\
&\approx \sgn\left(a\right)\sgn\left(b\right)\min\left(\left|a\right|,\left|b\right|\right) \text{,} \label{eq:SCFfunction} \\
g\left(a,b,s\right) &= b + \left(1-2s\right)a \label{eq:SCGfunction} \text{,}
\end{align}
and $\oplus$ is the bitwise XOR operation. SC decoding is initialized by setting $\alpha_{i,n} = y_i$ and the decoding schedule is such that the bits are decoded one by one from $u_0$ to $u_{N-1}$. Therefore, the function $f$ is given priority in SC decoding. At layer $0$, the elements of $\mathbf{u}$ are estimated as 
\begin{equation}
\hat{u}_i =
  \begin{cases}
    0 \text{,} & \text{if } i \in \mathcal{A}^c \text{ or } \alpha_{i,0} \geq 0\text{,}\\
    1 \text{,} & \text{otherwise.}
  \end{cases} \label{eq:SCestimate}
\end{equation}

\subsubsection{Successive-Cancellation List Decoding}

SCL decoding improves the error-correction performance of SC decoding by running multiple SC decoders in parallel. Unlike SC, instead of using (\ref{eq:SCestimate}) to estimate $\mathbf{u}$, each bit is estimated considering both its possible values $0$ and $1$. Therefore, at each bit estimation, the number of candidates doubles. In order to limit the exponential growth in the complexity of SCL, at each bit estimation a set of $L$ candidates are allowed to survive based on a path metric which is calculated as \cite{Alexios_LLR_SCLD,hashemi_SSCL_TCASI}
\begin{align} 
\PM_{i_\ell} &= \sum_{j = 0}^i \ln\left(1+\mathrm{e}^{-(1-2\hat{u}_{j_\ell})\alpha_{{j,0}_\ell}}\right) \text{,} \label{eq5} \\
&\approx \frac{1}{2}\sum_{j = 0}^{i}\sgn(\alpha_{{{j,0}_\ell}})\alpha_{{{j,0}_\ell}} - (1-2\hat{u}_{j_\ell})\alpha_{{{j,0}_\ell}} \text{,}
\end{align}
where $\ell$ is the path index and $\hat{u}_{j_\ell}$ is the estimate of bit $j$ at path $\ell$.

\subsubsection{Factor Graph Permutations}
\label{sec:polar:PGBPD}


Factor graph permutations are a way to provide multiple representations of a single code. It was observed in \cite{Kor09thesis} that there exists $n!$ different ways to represent a polar code by permuting the layers in its factor graph. In \cite{Doan_GLOBECOM}, the permutations in the factor graph layers were translated into permutations in the bit indices.
Each permutation is then decoded with a decoding algorithm and the codeword resulted from the decoding of the permutation with the best metric is selected as the output.

\section{Successive Factor Graph Permutations}

In this section, we first show that the number of permutations in the factor graph representation of polar and RM codes is higher than $n!$. Based on this observation, we introduce the SP scheme.

\begin{figure}[t]
  \centering
  \includegraphics[width=1\linewidth]{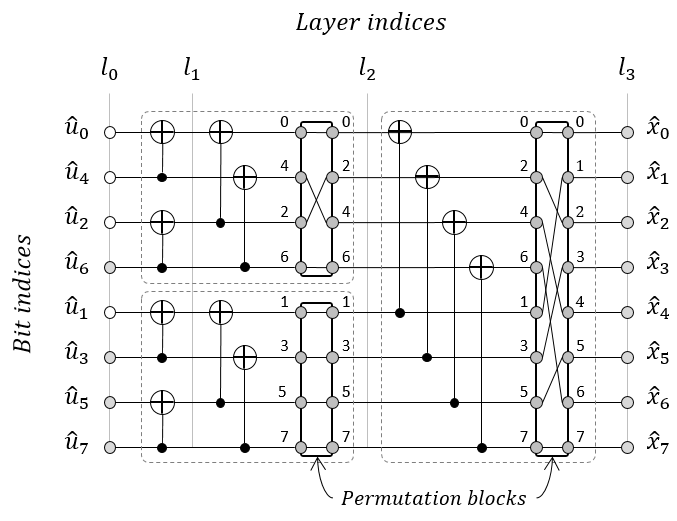}
  \caption{Permutations on a factor graph representation of a polar (RM) code with $N=8$, $K=4$, and $\{u_0,u_1,u_2,u_4\}\in \mathcal{A}^c$.}
  \label{fig:BPDecPerm}
\end{figure}

Consider the setting introduced in \cite{Doan_GLOBECOM} in which the bit indices are permuted instead of the layers in the factor graph representation. Let $\mathbf{b}_i = \{b_{{n-1}_i},\ldots,b_{0_i}\}$ be the binary expansion of the index $i$ and let $\pi_n : \{0,\ldots,n-1\} \rightarrow \{0,\ldots,n-1\}$ be a permutation of $n$ elements. We show a permutation in the bits of $\mathbf{b}_i$ as $\mathbf{b}_{\pi_n(i)} = \{b_{{\pi_n(n-1)}_i},\ldots,b_{\pi_n(0)_i}\}$. Clearly, since each index can be represented by $n$ bits, there are $n!$ permutations available. However, polar and RM codes can be represented as the concatenation of polar and RM sub-codes and it was shown in \cite{hashemi_TCOM} that each sub-code can be decoded independently. We use this idea to permute each sub-code independently. This is depicted for a polar (RM) code with $N=8$ and $K=4$ in Fig.~\ref{fig:BPDecPerm}. At each layer $l$ in the factor graph representation of polar and RM codes, there are $2^{n-l}$ polar or RM codes of length $2^{l}$. Therefore, the total number of permutations for a polar or RM code can be calculated as
\begin{equation}
\prod_{l = 0}^{n-1} \left(n-l\right)^{\left(2^{l}\right)} \text{,} \label{eq:permsAll}
\end{equation}
which quickly grows with $n$. For example, for a code of length $32$, (\ref{eq:permsAll}) results in $1658880$, while for a code of length $128$, the result is more than $1.9\times 10^{27}$. Since the memory requirements and the computational complexity of permutation decoding grows linearly with the number of permutations over which the decoding is performed, decoding over all of the above permutations is impossible.

It should be noted that if all the sub-codes in layer $l$ are permuted similarly, then the total number of permutations reduces to $n!$. In fact, one can represent the permutations in the bit indices as a nested permutation of cyclic shifts at each layer. At layer $l$ in the factor graph, there are $l$ cyclic shift permutations, therefore, the total number of permutations can be calculated as (\ref{eq:permsAll}). In order to limit the complexity of running multiple decoding attempts on different permutations, we devise a method to induce the best permutations during the course of decoding on the fly.

\subsection{Successive Permutation for SC Decoding}

SC decoding is serial in nature where the bits are decoded one by one. The success of SC decoding relies on the correct estimation of bits in the early stages of the decoding process. This is due to the fact that if a bit is estimated incorrectly, there is no way to correct it. The decoding process follows the calculation of LLR values at each decoding layer using $f$ and $g$ functions as in (\ref{eq:SCFfunction}) and (\ref{eq:SCGfunction}), with priority given to $f$ function. Therefore, it is crucial for SC decoding that the $f$ function provides reliable LLR values for the intermediate layers. Note that the reliability of a LLR value can be measured by taking the absolute value of it \cite{Ali_FSSCL}. Therefore, we can measure the reliability of a vector of LLR values $\bm{\alpha}$ of length $N$ as
\begin{equation}
\mathcal{R}(\bm{\alpha}) = \sum_{i = 0}^{N-1} \left|\alpha_i\right| \text{.}
\end{equation}
A larger $\mathcal{R}(\bm{\alpha})$ indicates a more reliable LLR vector.

In order to find the most reliable LLR vector resulting from the permutations, we successively permute the layers, considering only the cyclic shifts. We then pick the permutation with which the $f$ function results in the most reliable LLR values. Let us consider the vector of LLR values resulting from the $f$ function at layer $l$ as $\bm{\alpha}_l^f = \{\alpha_{0,l},\alpha_{1,l},\ldots,\alpha_{2^l-1,l}\}$. The reliability of $\bm{\alpha}_l^f$ can be computed as
\begin{equation}
\mathcal{R}(\bm{\alpha}_l^f) = \sum_{i = 0}^{2^l-1} \left|\alpha_{i,l}\right| = \sum_{i = 0}^{2^l-1} \left|f\left(\alpha_{i,l+1},\alpha_{i+2^l,l+1}\right)\right| \text{.}
\end{equation}
Now consider all the cyclic shift permutations $\pi_{l+1}$ at layer $l+1$. We find the permutation $P_{l+1}$ at layer $l+1$ which results in the largest $\mathcal{R}(\bm{\alpha}_l^f)$ in accordance with
\begin{equation}
P_{l+1} \!=\! \argmax_{p_{l+1} \in \pi_{l+1}} \sum_{i = 0}^{2^l-1} \left|f\!\left(\alpha_{p_{l+1}(i),l+1},\alpha_{p_{l+1}(i+2^l),l+1}\right)\right| \text{.} \label{eq:bestPerm}
\end{equation}

The process is successively repeated at each layer whenever the $f$ function calculation is performed. The permutation is then fixed at that layer and is used whenever the corresponding $g$ function calculation is conducted. Note that at each layer $l$, there are only $l$ cyclic shift permutations to be checked in (\ref{eq:bestPerm}). Since only one permutation is selected at each layer, SP algorithm does not change the memory requirements of SC. However, it will add to the computational complexity of SC by a factor of $n$, in order to perform (\ref{eq:bestPerm}). We now analyse the effect of SP for SC decoding in case of RM and polar codes.

\subsubsection{RM Codes}

The rate of RM codes is limited to those in (\ref{eq:RMRate}). In fact, (\ref{eq:RMRate}) guarantees that all of the rows in $\mathbf{G}^{\otimes n}$ which have the same Hamming weight are either included in the information set $\mathcal{A}$ or the frozen set $\mathcal{A}^c$. On the other hand, the permutations in each layer only permutes the bit indices which correspond to the same Hamming weight. Therefore in RM codes, permutations do not change the location of information and frozen bits as seen by the SC decoder. In other words, while the order of the bits that are decoded by SC is changed, the frozen/information bit pattern remains unaltered. This specifically results in significant improvement in error-correction performance when SP is used with SC decoding on RM codes. This is due to the fact that while the code remains unchanged, the bits which are more likely to be decoded correctly are decoded in the earlier stages of the SC decoding process.

\subsubsection{Polar Codes}

Unlike RM codes, polar codes are constructed based on reliabilities of bit indices under SC decoding. In other words, the frozen/information bit pattern of polar codes is optimized to result in the best error-correction performance under SC decoding for a specific channel condition. Since the bit indices of information bits are not selected based on their Hamming weights, the permutations may result in a different frozen/information bit pattern which may not be optimized for the specific channel on which the transmission takes place. This in turn may result in error-correction performance loss when SP is used along with SC. In order to resolve this issue, we only apply SP on the sub-codes of polar codes which are RM codes. For decoding the rest of the code, we use the conventional SC decoding. This will guarantee that the frozen/information bit pattern remains unchanged in case of polar codes.

\subsection{Successive Permutation for SCL decoding}

SCL decoding with list size $L$ can be considered as $L$ different SC decoders which are running in parallel and at each bit estimation, $L$ candidates are allowed to survive. In order to use SP with SCL decoding, we apply SP to each of the $L$ SC decoders in a SCL decoder. More formally, consider the vector of LLR values resulting from the $f$ function at layer $l$ for the $\ell$-th path in the SCL decoder as $\bm{\alpha}_{l_\ell}^f = \{\alpha_{{0,l}_\ell},\alpha_{{1,l}_\ell},\ldots,\alpha_{{2^l-1,l}_\ell}\}$. For each path $\ell$, we find the permutation $P_{{l+1}_\ell}$ at layer $l+1$ which results in the largest $\mathcal{R}(\bm{\alpha}_{l_\ell}^f)$ in accordance with
\begin{equation}
P_{{l+1}_\ell} \!=\! \argmax_{p_{{l+1}_\ell} \in \pi_{{l+1}_\ell}} \sum_{i = 0}^{2^l-1} \left|f\!\left(\alpha_{{p_{l+1}(i),l+1}_\ell},\alpha_{{p_{l+1}(i+2^l),l+1}_\ell}\right)\right| \text{,} \label{eq:bestPermSCL}
\end{equation}
where $\pi_{{l+1}_\ell}$ represents all the cyclic shift permutations for path $\ell$ at layer $l+1$. Consider after SP, bit $i$ at path $\ell$ is permuted to a new position $\pi^\ell_n(i)$. The path metric associated with the $\ell$-th path when SP is applied to SCL decoding can be calculated as
\begin{align} 
\PM_{i_\ell} &\!=\! \sum_{j = 0}^i \ln\left(1+\mathrm{e}^{-(1-2\hat{u}_{\pi^\ell_n(j)})\alpha_{\pi^\ell_n(j),0}}\right) \text{,} \label{eq5} \\
&\!\approx\! \frac{1}{2}\sum_{j = 0}^{i}\sgn(\alpha_{\pi^\ell_n(j),0})\alpha_{\pi^\ell_n(j),0} - (1-2\hat{u}_{\pi^\ell_n(j)})\alpha_{\pi^\ell_n(j),0} \text{,}
\end{align}
where $\ell$ is the path index and $\hat{u}_{j_\ell}$ is the estimate of bit $j$ at path $\ell$.

We will use the same considerations when using SP for SCL decoding in case of polar and RM codes, i.e., since permutations do not change the frozen/information bit pattern in RM codes, we directly use SP in SCL decoding for RM codes and only apply SP on the sub-codes of polar codes which are RM codes.

\section{Results}

In this section, we evaluate the application of SP on SC and SCL decoding for RM and polar codes. The simulations in this section are carried out for $\mathcal{RM}(128,64)$ and $\mathcal{P}(128,64)$, where the polar code is constructed for SNR~$= 6$~dB. The results are reported based on FER. The SC decoder with SP is denoted as SPSC and the SCL decoder with SP is denoted as SPSCL. SCL($L$) and SPSCL($L$) denote the SCL and SPSCL decoders with list size $L$ respectively.

\subsection{RM Codes}

\begin{figure}
  \centering
  \begin{tikzpicture}
  \pgfplotsset{
    label style = {font=\fontsize{9pt}{7.2}\selectfont},
    tick label style = {font=\fontsize{7pt}{7.2}\selectfont}
  }

\begin{axis}[
	scale = 1,
    ymode=log,
    xlabel={$E_b/N_0$ [\text{dB}]}, xlabel style={yshift=0.5em},
    xtick={2,2.5,3,3.5,4,4.5,5,5.5,6},
    ylabel={FER}, 
    grid=both,
    ymajorgrids=true,
    xmajorgrids=true,
    grid style=dashed,
    width=\columnwidth, height=6.75cm,
    thick,
    mark size=3,
    legend style={
      anchor={center},
      cells={anchor=west},
      column sep= 2mm,
      font=\fontsize{7pt}{7.2}\selectfont,
    },
    legend to name=perf-legend-RM128,
    legend columns=4,
]

\addplot[
    color=blue,
    mark=o,
    thick,
    mark size=3,
    dashed,
]
table {
2 0.3944
2.5 0.2477
3 0.1311
3.5 0.06
4 0.0246
4.5 0.00705119
5 0.00135142
5.5 0.000228076
6 2.55069e-05
};
\addlegendentry{SC}

\addplot[
    color=black,
    mark=square,
    thick,
    mark size=3,
    dashed,
]
table {
2 0.2148
2.5 0.1038
3 0.0425
3.5 0.0139
4 0.00285747
4.5 0.000455436
5 5.86839e-05
5.5 3.16132e-06
};
\addlegendentry{SCL($2$)}

\addplot[
    color=red,
    mark=triangle,
    thick,
    mark size=3,
    dashed,
]
table {
2 0.1174
2.5 0.0455
3 0.015
3.5 0.00266489
4 0.000448867
4.5 4.13626e-05
5 2.57645e-06
};
\addlegendentry{SCL($4$)}

\addplot[
    color=brown,
    mark=star,
    thick,
    mark size=3,
    dashed,
]
table {
2 0.0667
2.5 0.0214
3 0.00519292
3.5 0.000914854
4 0.000111186
4.5 6.8569e-06
};
\addlegendentry{SCL($8$)}

\addplot[
    color=blue,
    mark=o,
    thick,
    mark size=3,
]
table {
2 0.2532
2.5 0.1369
3 0.0596
3.5 0.0216
4 0.0067714
4.5 0.00139978
5 0.000209307
5.5 2.01349e-05
};
\addlegendentry{SPSC}

\addplot[
    color=black,
    mark=square,
    thick,
    mark size=3,
]
table {
2 0.1224
2.5 0.0481
3 0.0142
3.5 0.00345197
4 0.00065463
4.5 6.93805e-05
5 6.78433e-06
};
\addlegendentry{SPSCL($2$)}

\addplot[
    color=red,
    mark=triangle,
    thick,
    mark size=3,
]
table {
2 0.0631
2.5 0.0194
3 0.0042555
3.5 0.000628615
4 8.62598e-05
4.5 7.23574e-06
};
\addlegendentry{SPSCL($4$)}

\addplot[
    color=brown,
    mark=star,
    thick,
    mark size=3,
]
table {
2 0.0363
2.5 0.00940911
3 0.00168169
3.5 0.000199533
4 2.19217e-05
};
\addlegendentry{SPSCL($8$)}

\addplot[
    color=magenta,
    mark=asterisk,
    thick,
    mark size=3,
    dashed,
]
table {
2 0.0401
2.5 0.0104
3 0.00193945
3.5 0.000250297
4 2.1333e-05
};
\addlegendentry{SCL($16$)}


\addplot[
    color=magenta,
    mark=asterisk,
    thick,
    mark size=3,
]
table {
2 0.0233
2.5 0.00575871
3 0.000895937
3.5 0.000126085
4 1.39234e-05
};
\addlegendentry{SPSCL($16$)}


\addplot[
    thick,
    dotted,
]
table {
2 0.0162
2.5 0.0037802
3 0.000641206
3.5 0.000109529
4 1.23583e-05
};
\addlegendentry{MAP Lower Bound\kern-4em}

\end{axis}
\end{tikzpicture}
  \\
  \ref{perf-legend-RM128}
  \caption{FER performance of $\mathcal{RM}(128,64)$ under SC, SPSC, SCL, and SPSCL decoding.}
  \label{fig:perfRM}
\end{figure}
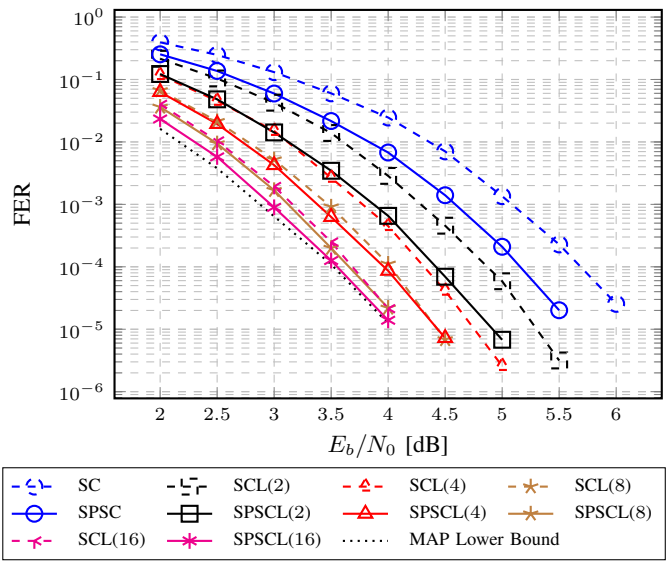

Fig.~\ref{fig:perfRM} shows the effect of applying SP on SC and SCL decoding when RM codes are used. It can be seen that SPSC outperforms SC by around $0.5$~dB at a target FER of $10^{-4}$. In addition, SP brings significant improvements when applied to SCL decoding. For example, SPSCL($4$) slightly outperforms SCL($8$) at a target FER of $10^{-4}$ and SPSCL($16$) provides an FER performance which is less than $0.05$~dB away from the MAP lower bound at FER of $10^{-4}$.

\subsection{Polar Codes}

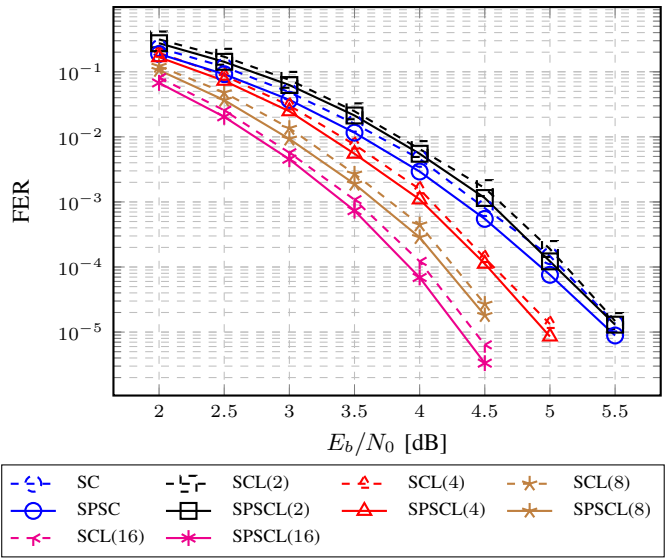
\begin{figure}
  \centering
  \begin{tikzpicture}
  \pgfplotsset{
    label style = {font=\fontsize{9pt}{7.2}\selectfont},
    tick label style = {font=\fontsize{7pt}{7.2}\selectfont}
  }

\begin{axis}[
	scale = 1,
    ymode=log,
    xlabel={$E_b/N_0$ [\text{dB}]}, xlabel style={yshift=0.5em},
    xtick={2,2.5,3,3.5,4,4.5,5,5.5},
    ylabel={FER}, 
    grid=both,
    ymajorgrids=true,
    xmajorgrids=true,
    grid style=dashed,
    width=\columnwidth, height=6.75cm,
    thick,
    mark size=3,
    legend style={
      anchor={center},
      cells={anchor=west},
      column sep= 2mm,
      font=\fontsize{7pt}{7.2}\selectfont,
    },
    legend to name=perf-legend-PC128,
    legend columns=4,
]

\addplot[
    color=blue,
    mark=o,
    thick,
    mark size=3,
    dashed,
]
table {
2	0.2316
2.5	0.1192
3	0.0486
3.5	0.0159
4	0.00458842
4.5	0.000801796
5	0.000147126
5.5	1.34881e-05
};
\addlegendentry{SC}

\addplot[
    color=black,
    mark=square,
    thick,
    mark size=3,
    dashed,
]
table {
2	3.14E-01
2.5	0.1702
3	7.52E-02
3.5	0.0247
4	0.0065
4.5	0.00163244
5	0.000189225
5.5	1.48E-05
};
\addlegendentry{SCL($2$)}

\addplot[
    color=red,
    mark=triangle,
    thick,
    mark size=3,
    dashed,
]
table {
2	0.1958
2.5	0.087
3	0.0303
3.5	0.0076
4	0.00156878
4.5	0.000137151
5	1.33E-05
};
\addlegendentry{SCL($4$)}

\addplot[
    color=brown,
    mark=star,
    thick,
    mark size=3,
    dashed,
]
table {
2	0.1222
2.5	0.0479
3	0.0136
3.5	0.00269818
4	4.47E-04
4.5	2.70E-05
};
\addlegendentry{SCL($8$)}

\addplot[
    color=blue,
    mark=o,
    thick,
    mark size=3,
]
table {
2	0.188
2.5	0.0919
3	0.0371
3.5	0.0116
4	0.00292929
4.5	0.000548342
5	7.58E-05
5.5	8.86E-06
};
\addlegendentry{SPSC}

\addplot[
    color=black,
    mark=square,
    thick,
    mark size=3,
]
table {
2	0.2748
2.5	0.1422
3	0.0626
3.5	0.0213
4	0.0055
4.5	0.001149
5	1.22E-04
5.5	1.30E-05
};
\addlegendentry{SPSCL($2$)}

\addplot[
    color=red,
    mark=triangle,
    thick,
    mark size=3,
]
table {
2	0.1673
2.5	0.0727
3	0.0247
3.5	0.0055
4	0.00109187
4.5	0.000112009
5	8.55E-06
};
\addlegendentry{SPSCL($4$)}

\addplot[
    color=brown,
    mark=star,
    thick,
    mark size=3,
]
table {
2	0.1057
2.5	0.0367
3	0.0093
3.5	0.00188693
4	0.000285316
4.5	1.81E-05
};
\addlegendentry{SPSCL($8$)}

\addplot[
    color=magenta,
    mark=asterisk,
    thick,
    mark size=3,
    dashed,
]
table {
2	0.0815
2.5	0.0256
3	0.0057
3.5	0.00108324
4	1.22E-04
4.5	6.37E-06
};
\addlegendentry{SCL($16$)}

\addplot[
    color=magenta,
    mark=asterisk,
    thick,
    mark size=3,
]
table {
2	6.74E-02
2.5	2.03E-02
3	4.53E-03
3.5	7.34E-04
4	6.90E-05
4.5	3.33E-06
};
\addlegendentry{SPSCL($16$)}

\end{axis}
\end{tikzpicture}
  \\
  \ref{perf-legend-PC128}
  \caption{FER performance of $\mathcal{P}(128,64)$ under SC, SPSC, SCL, and SPSCL decoding. The polar code is constructed for SNR~$=6$~dB and an $11$-bit CRC is used for SCL and SPSCL.}
  \label{fig:perfPC}
\end{figure}

Fig.~\ref{fig:perfPC} illustrates the effect of applying SP on SC and SCL decoding when polar codes are used. An $11$-bit cyclic redundancy check (CRC) is used in case of SCL and SPSCL decoders. It can be seen that the FER performance improvement brought by applying SP is around $0.1$~dB at a target FER of $10^{-4}$. This improvement is smaller than the case of RM codes since we only apply SP on the sub-codes of polar codes which are RM codes. In fact for polar codes, SP is more beneficial when the polar code has many RM sub-codes. Therefore, the performance of SP for polar codes is dependent on the polar codes construction.

\section{Conclusion} \label{sec:conc}

In this paper, we first derived the total number of factor graph permutations for Reed-Muller (RM) and polar codes. We then proposed successive factor graph permutations for RM and polar codes which can find the suitable factor graph permutations on the fly. We showed that the proposed successive permutation (SP) scheme significantly improves the error-correction performance of successive-cancellation (SC) and SC list (SCL) decoders. Our results for RM and polar codes of length $128$ and rate $0.5$ show that by using SP, the error-correction performance of RM and polar codes under SC and SCL decoding can be improved by up to $0.5$~dB and $0.1$~dB respectively, at a target frame error rate of $10^{-4}$. Moreover, SP enables RM code of length $128$ and rate $0.5$ to approach within $0.05$~dB of the maximum a posteriori (MAP) lower bound with SCL decoder of list size $16$.


\end{document}